\documentclass[10pt,twocolumn,journal]{IEEEtran}
\usepackage{cite}
\usepackage{amsmath,amssymb,amsfonts}
\usepackage{algorithmic}
\usepackage{algorithm}
\usepackage{graphicx}
\usepackage{textcomp}
\usepackage{mathtools}
\usepackage{subcaption}
\usepackage{enumitem}
\usepackage{bm}
\usepackage{bbm}
\usepackage{tabularx}
\setdescription{noitemsep,topsep=3pt,parsep=0pt,partopsep=0pt,font=\bfseries}
\setenumerate{noitemsep,topsep=3pt,parsep=0pt,partopsep=0pt}
\usepackage{amsthm}

\usepackage[table]{xcolor}
\usepackage{booktabs}
\usepackage{multirow}
\usepackage{makecell}

\renewcommand{\thetable}{\Roman{table}} 
\makeatletter
\renewcommand{\p@subtable}{}                       
\makeatother

\makeatletter
\renewcommand{\p@subfigure}{}                       
\makeatother

\theoremstyle{plain}

\theoremstyle{definition}

\def\BibTeX{{\rm B\kern-.05em{\sc i\kern-.025em b}\kern-.08em
    T\kern-.1667em\lower.7ex\hbox{E}\kern-.125emX}}

\begin{document}

\title{Joint Alignment and Denoising for Event-Based Vision Sensors Using Regret-based Pareto Optimization}

\author{\IEEEauthorblockN{Shimpei Harada, Junya Hara,~\IEEEmembership{Member, IEEE}, Hiroshi Higashi~\IEEEmembership{Member, IEEE}, and Yuichi Tanaka,~\IEEEmembership{Senior Member, IEEE}}\\
\thanks{This work was supported in part by JSPS KAKENHI under Grant 23H01415, 22H05163, and 22K12500, JST AdCORP under Grant JPMJKB2307.

 S. Harada, J. Hara, H. Higashi and Y. Tanaka are with the Graduate School of Engineering, The University of Osaka, Osaka, Japan.}}

\markboth{Journal of \LaTeX\ Class Files,~Vol.~18, No.~9, September~2020}%
{How to Use the IEEEtran \LaTeX \ Templates}

\maketitle

\begin{abstract}
This paper proposes a joint alignment and denoising method for event-based vision sensors (EVSs). 
Existing signal processing methods for EVSs typically perform event alignment (EA) and event denoising (ED) as separate modules. However, this separation creates a dilemma: without ED, EA is biased by noise, whereas without EA, ED struggles to distinguish signal events from noise ones. To address this dilemma, we jointly optimize EA and ED by formulating a bi-objective Pareto optimization problem. Our formulation is built upon a contrast map that counts the number of events localized in each pixel. With a contrast map, we can formulate EA as maximizing its variance and ED as minimizing the variance. We cast these two conflicting problems as a Pareto optimization and use a regret strategy to obtain a solution. Experimental results on denoising and motion estimation demonstrate that our method achieves improvements against alternative ones.

\end{abstract}


\section{Introduction}\label{sec:intro}

\IEEEPARstart{E}{vent}‑based vision sensors (EVSs) are an emerging class of vision sensors \cite{gallegoEventbasedVisionSurvey2020}, where events are triggered by local brightness changes in a scene. EVSs yield asynchronous events at each pixel with microsecond latency and ultra‑high dynamic range, forming a 3-D (2-D pixel coordinates and time) event stream. Compared with high-speed RGB cameras, EVSs can be deployed at a lower cost and are attractive for many real-time imaging applications \cite{gallegoEventbasedVisionSurvey2020}.

\begin{figure}[t]
\centering
\includegraphics[width = \linewidth]{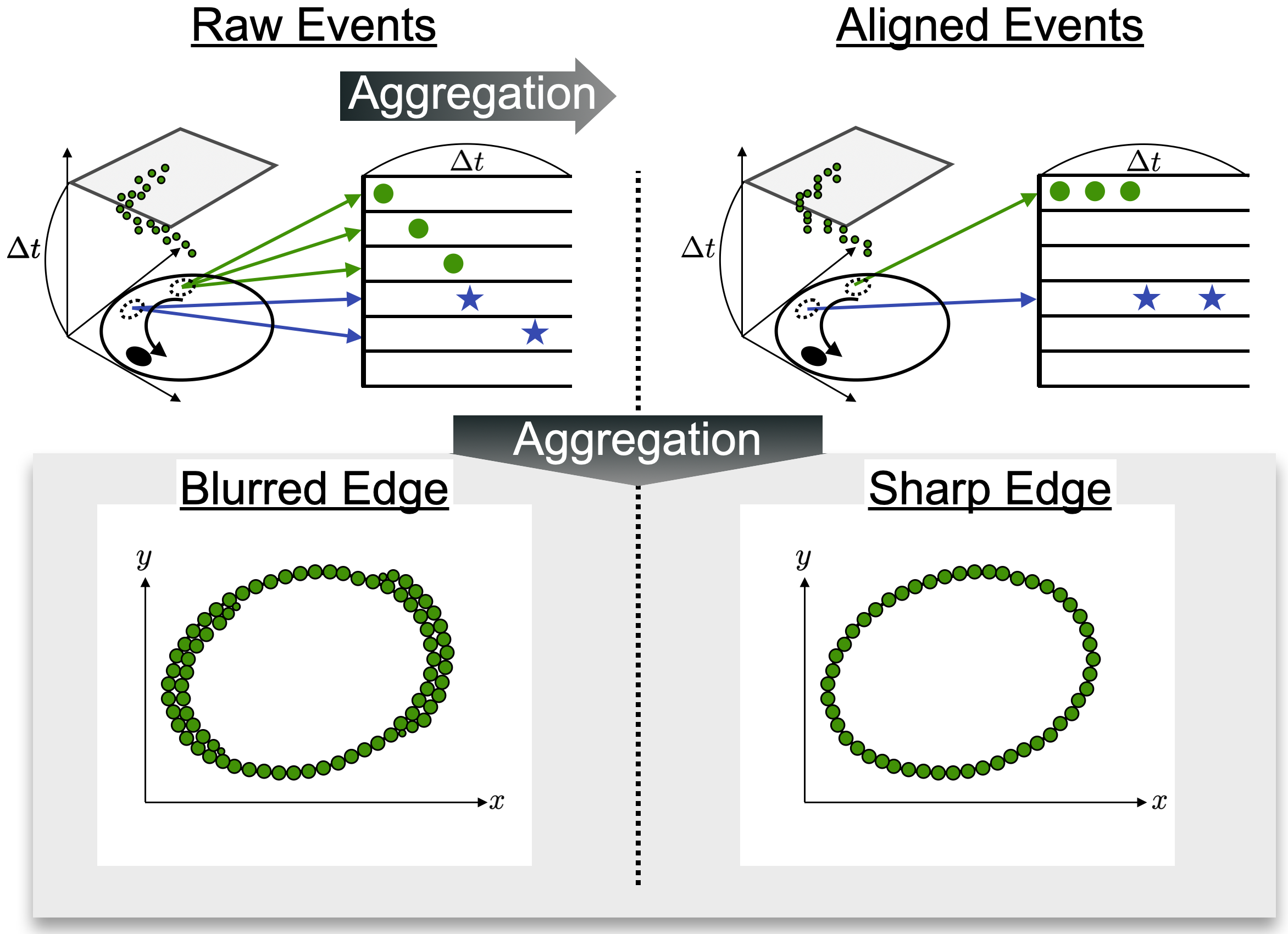} 
\caption{
Top: Illustration of event drift and alignment of a rotating dot. In the raw event stream, events corresponding to the same point on the moving object (shown as green circles or blue stars) are displaced to multiple pixel locations over time. EA compensates for these displacements.
Bottom: Example of aggregated events. Raw (drifted) events form blurred object edges, whereas aligned events show sharp edges.}
\label{fig:intro}
\end{figure}

Under fixed light intensity and sensor position, EVSs detect events in response to moving objects in a scene. 
These events often concentrate along object edges since brightness changes in moving objects occur primarily at their boundaries. As a result, event coordinates drift over time along the direction of motion: We usually observe blurred object edges when events are aggregated within a short time window, as visualized in Fig.~\ref{fig:intro}.
Event drifts may degrade boundary localization of objects and complicate boundary matching across windows, which in turn negatively affects downstream applications such as tracking, segmentation, and motion estimation \cite{shibaEventCollapseContrast2022}.

Event alignment (EA) performs motion compensation for drifted events as shown in Fig.~\ref{fig:intro} \cite{gallegoEventbasedVisionSurvey2020, gallegoUnifyingContrastMaximization2018}. EA aims to estimate motion vectors (MVs) for each event and compensate their coordinates according to the estimated MVs.
However, most EA approaches implicitly assume noise-free event streams, and MVs are calculated even for noise events that do not carry actual motion information. This results in a biased EA, causing the disappearance or duplication of object edges.

To avoid misalignments in EA, many methods apply outlier removal and/or denoising to raw event streams as a preprocessing \cite{delbruckFramefreeDynamicDigital2008,fengEventDensityBased2020a,zhangNeuromorphicImagingDensitybased2023}. 
Event denoising (ED) modules operate on the simple assumption that noise is sparse in the 3-D space. However, drifted events can also appear sparse. In other words, ED modules are primarily designed for already-aligned event streams. Thus, ED without EA beforehand often removes signal (but drifted) events. 

Here, we face a ``chicken-and-egg" dilemma: 
\textit{Without ED, EA is biased, and without EA, ED overestimates noise events}.
A recent method in \cite{shibaSimultaneousMotionNoise2025} alternates EA and ED modules over iterations to alleviate this dilemma.
However, in this method, EA and ED are still designed as separate modules, thereby depending on manual hyperparameter tuning to control the trade-off between them. For example, an aggressive ED setting removes many drifted signal events and causes object edges to disappear, whereas a conservative ED setting leaves noise events that lead to biased EA.

The trade-off between EA and ED can be quantified with the contrast map. The contrast map is a 2-D image obtained by counting the number of events localized in each pixel. This map is widely utilized for EA \cite{gallegoUnifyingContrastMaximization2018, shiba2024secrets}. On the contrast map, EA concentrates events on a small subset of pixels, yielding high-contrast edges, whereas ED tends to reduce the overall contrast level by removing noise events.
Based on these contrast behaviors, EA and ED can be formulated as follows:
\begin{description}
    \item[EA] is formulated as compensating the drift so as to \textit{maximize} the variance of the contrast map. 
    \item[ED] is formulated as removing events so as to \textit{minimize} the variance of the contrast map. 
\end{description}


Our main idea is to optimize the trade-off between EA and ED by formulating the problem as a bi-objective \textit{Pareto optimization}. This formulation provides a set of trade-off solutions between EA and ED. To obtain a well-compromised solution, we employ a \textit{regret} strategy \cite{savage1951theory, wierzbicki1980use}: It aims to minimize the worst-case difference from the best possible values (i.e., baselines) on any objective functions. With this strategy, the trade-off between EA and ED can be controlled inside the optimization as baselines, rather than sensitive hyperparameter tuning. We can solve it with a gradient-based solver.

To validate the effectiveness of our method, we perform denoising and motion estimation experiments on synthetic and real-world event datasets. Numerical results demonstrate that the proposed method achieves improvements against alternative ones in both experiments.

\textit{Notation:} 
A vector and a matrix are denoted by boldface lower and upper cases, respectively.
We denote a $\ell_1$ norm and Frobenius norm by $\|\cdot\|_1$ and $\|\cdot\|_{\text{F}}$, respectively. The cardinality of a set $\mathcal{A}$ is denoted by $|\mathcal{A}|$.
The element-wise product is denoted by $\circ$.

\section{Related Work}\label{sec:related}

In this section, we first introduce the measurement model of EVSs. Then, we briefly review existing studies on EA and ED, as well as recent approaches that jointly address both tasks.

\subsection{Event Generation Model}
EVSs asynchronously and independently detect brightness changes at each pixel $\mathbf{x}\in\mathbb{R}^2$ \cite{lichtsteiner128Backslashtimes1282008}.
Let $I(\mathbf{x},t)$ be the brightness at $\mathbf{x}$ on time instance $t$.
An EVS measures the brightness change by the temporal log-difference as follows:
\begin{equation}
\Delta\ln I(\mathbf{x},t)=\ln I(\mathbf{x},t)-\ln I(\mathbf{x},t-\Delta t).
\end{equation}
The $k$th event $e_k\coloneqq(\mathbf{x}_k, t_k, p_k)$ is triggered if $p_k\Delta\ln I(\mathbf{x}_k,t_k) \geq C$, where $p_k\in \{-1, 1\}$ is the polarity, indicating whether the brightness is increased ($p_k=+1$) or decreased ($p_k=-1$), and $C>0$ is a controllable contrast threshold. 
A lower $C$ allows the sensor to detect smaller brightness changes, thereby capturing finer motion details, at the cost of a higher rate of noise events.

\subsection{Event Alignment Methods}

As mentioned, events often drift along the direction of motion. To compensate for this drift, EA methods aim to estimate event-wise MVs. These methods can be categorized into the following two approaches:
\begin{description}
    \item[3-D point-based approach:] This approach treats event streams as 3-D point clouds. In this approach, an event stream is first split into two short, consecutive time windows. MVs are then estimated so that one point cloud best matches the other \cite{nunes2020entropy, nunes2021live, nunes2021robust}. 
    \item[2-D image-based approach:] This approach is based on the intuition that the accumulation of well-aligned events onto a 2-D image plane (contrast map) should contain sharp object edges. In this approach, MVs are estimated by optimizing a sharpness measure of the contrast map (e.g., its variance or entropy). \cite{gallegoUnifyingContrastMaximization2018,shiba2024secrets}.

\end{description}

Note that these EA approaches are implicitly designed for noise-free event streams: their performance is limited under noisy conditions. Thus, existing EA methods result in the disappearance or duplication of object edges without prior ED.

\subsection{Event Denoising Methods}

The acquisition process of EVSs usually yields numerous noise events, that are independent of object motion.
This noise is typically distributed randomly in the 3-D space.
For ED, two approaches have mainly been proposed.
\begin{description}
    \item[Density-based approach:] This approach is based on the assumption that signal events are more densely distributed than noise events in spatiotemporal space \cite{delbruckFramefreeDynamicDigital2008,fengEventDensityBased2020a,zhangNeuromorphicImagingDensitybased2023, padalaNoiseFilteringAlgorithm2018, liuDesignSpatiotemporalCorrelation2015, khodamoradiNspaceSpatiotemporalFilter2018, guoLowCostLatency2022, wuProbabilisticUndirectedGraph2020, guoHashHeatComplexityHashingbased2020, xuEffectiveTargetBinarization2016}. To implement this, these methods remove events with few neighbors within a small radius sphere in the 3-D space.
    \item[Learning-based approach:] Learning-based denoising approaches using deep neural networks have achieved good performance if enough amount of training data is available \cite{baldwinEventProbabilityMask2020, duanEventZoomLearningDenoise2021, xieDVSImageNoise2018, guoLowCostLatency2022, afsharEventbasedFeatureExtraction2020, liEventStreamSuperresolution2021}. Training data is usually annotated using density-based approaches \cite{guoLowCostLatency2022}.
\end{description}

These ED approaches assume a static scene where signal events are aligned in time. However, in most practical settings, EVSs observe dynamic scenes and signal events drift over time. Thus, existing ED modules often misclassify signal events as noise events without prior EA.

\subsection{Joint Optimization Method}\label{subsec:joint}
To address EA and ED jointly, an alternating optimization scheme is proposed in \cite{shibaSimultaneousMotionNoise2025}. This approach iterates the following two steps: 1) EA step that maximizes the variance of a contrast map, and 2) ED step that classifies events via a hard thresholding on the contrast map.

Although this approach considers the chicken-and-egg dilemma between EA and ED, it still treats them as separate modules. Due to this separation, careful hyperparameter tuning is required to control the trade-off between EA and ED. Moreover, since ED is performed as hard thresholding, drifted events that are misclassified as noise in the early ED step cannot be compensated in later EA steps. As a result, the fundamental dilemma between EA and ED still remains.

\section{Joint Alignment and Denoising}\label{sec:proposed}

\begin{figure}[t]
\centering
\includegraphics[width = \linewidth]{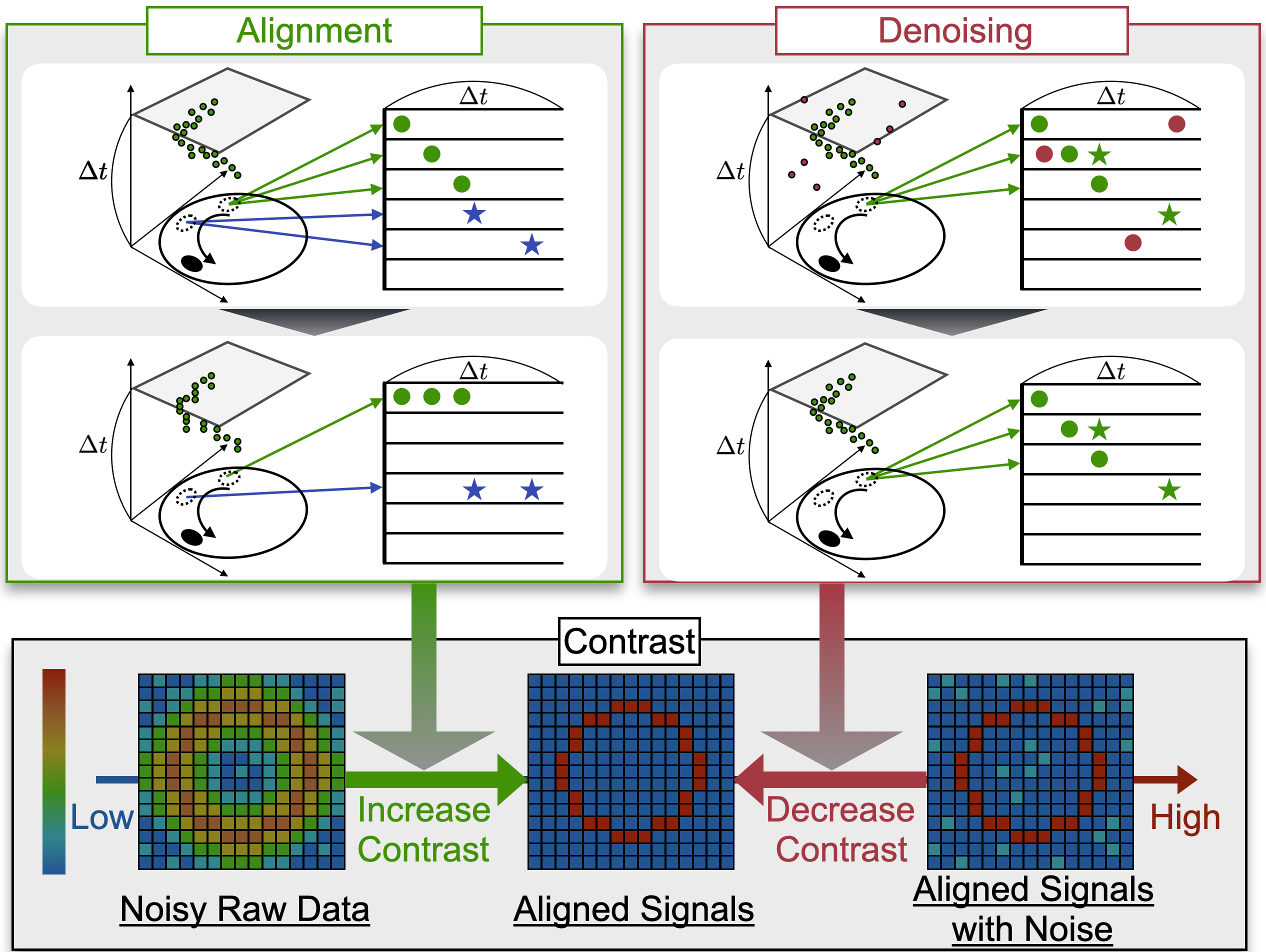} 
\caption{
The trade-off between alignment and denoising. Alignment seeks to increase the contrast, whereas denoising acts to decrease it.}
\label{fig:idea_method}
\end{figure}

In this section, we formulate a joint EA and ED. 
Our formulation is built upon three components.
\begin{enumerate}
    \item \textit{Contrast map}:
    We first introduce a contrast map \cite{gallegoUnifyingContrastMaximization2018,shiba2024secrets}, which is widely utilized for EA. This is obtained by counting the number of events localized in each pixel.
    \item \textit{Formulation of EA and ED using contrast map}:
    Next, we formulate both EA and ED using the variance of the contrast map (VCM), respectively. Similar to \cite{gallegoUnifyingContrastMaximization2018,shiba2024secrets}, EA is formulated as the maximization of the VCM. We then newly formulate ED as the minimization of the VCM.
    \item \textit{Joint EA and ED as Pareto optimization}:
    We cast the joint EA and ED as a regret-based bi-objective Pareto optimization.
\end{enumerate}
In the following, we describe the building blocks in detail.

\subsection{Contrast Map}
Here, we first introduce the contrast map.
Let $\mathcal{E}=\{e_k\}_{k=1}^{N_e}$ be an observed event stream, where $N_e$ is the number of events.
Let $\Omega \coloneqq [0,W)\times[0,H)\subset\mathbb{R}^2$ denote the image plane, and let
$\Omega_{ij}\coloneqq [j,j+1)\times[i,i+1)$ for $i=0,\ldots,H-1$ and $j=0,\ldots,W-1$ be the $(i,j)$th pixel cell.

The contrast map $\mathbf{M}(\mathcal{E}):\mathcal{E}\to\mathbb{R}^{H\times W}$ is a mapping whose $(i,j)$th entry is given by
\begin{equation}\label{eq:contrast_map}
    \begin{split}
        \bigl[\mathbf{M}(\mathcal{E})\bigr]_{ij}&\coloneqq\int_{\Omega_{ij}}\sum_{k=1}^{N_e}\delta\left(\mathbf{x}-\mathbf{x}_k\right)\mathrm{d}\mathbf{x}\\
        &=\sum_{k=1}^{N_e}\mathbb{I}\bigl(\mathbf{x}_k \in \Omega_{ij}\bigr),
    \end{split}
\end{equation}
where $\delta(\cdot)$ is the Dirac delta and $\mathbb{I}(\cdot)$ denotes the indicator function. 
According to \eqref{eq:contrast_map}, each pixel value in the contrast map stores the number of events that fall within $\Omega_{ij}$.

\subsection{Formulation of EA and ED Using Contrast Map}

We formulate EA and ED using the contrast map.
\subsubsection{EA Formulation}
Let $T(\mathbf{x}, t, \boldsymbol{\theta})$ be a transformation function induced by an assumed motion model with motion parameters $\boldsymbol{\theta}$. Given an observed event $e_k=(\mathbf{x}_k,\,t_k,\,p_k)$ and a fixed reference time $t_{\text{ref}}$, the function $T(\mathbf{x}_k,t_k, \bm{\theta})$ maps the location of $e_k$ from $\mathbf{x}_k$ to the motion-compensated location $\mathbf{x}^\prime_k(\bm{\theta})$ where the same physical point on a object would be observed at $t_{\mathrm{ref}}$, i.e., $\mathbf{x}^\prime_k(\bm{\theta})= T(\mathbf{x}_k,t_k, \bm{\theta})$\footnote{For example, under the translation transform model, $\mathbf{x}^\prime_k(\bm{\theta})= \mathbf{x}_k+(t_k-t_{\mathrm{ref}})\bm{\theta}$, where $\bm{\theta}=[v_x,v_y]^\top$.}.
This transformation yields a collection of aligned events $\mathcal{E}^\prime(\bm{\theta})=\bigl\{\,(\mathbf{x}^\prime_k(\bm{\theta}),\,t_k,\,p_k)\,\bigr\}^{N_e}_{k=1}$.

Applying \eqref{eq:contrast_map} to $\mathcal{E}^\prime(\bm{\theta})$, we can obtain $\mathbf{M}\left(\mathcal{E}^\prime(\bm{\theta})\right)$\footnote{The contrast map in \eqref{eq:iwe} is commonly referred to as the image of warped events (IWE).} whose $(i,j)$th entry is
\begin{equation}\label{eq:iwe}
    \bigl[\mathbf{M}\left(\mathcal{E}^\prime(\bm{\theta})\right)\bigr]_{ij}=\sum_{k=1}^{N_e}\mathbb{I}\bigl(\mathbf{x}'_k(\boldsymbol{\theta}) \in \Omega_{ij}\bigr).
\end{equation}
When events are correctly aligned, they accumulate along object edges on $\mathbf{M}\left(\mathcal{E}^\prime(\bm{\theta})\right)$: only a small subset of pixels take large values, while most pixels remain zero. 

Similar to \cite{gallegoUnifyingContrastMaximization2018,shiba2024secrets}, we employ the variance of $\mathbf{M}\left(\mathcal{E}^\prime(\bm{\theta})\right)$ to evaluate how well events are aligned:
\begin{equation}\label{eq:ea_objective}
    f_\text{EA}(\bm{\theta})=\frac{1}{HW}\sum_{i=0}^{H-1}\sum_{j=0}^{W-1}\left(\bigl[\mathbf{M}\left(\mathcal{E}^\prime(\bm{\theta})\right)\bigr]_{ij}-\mu_m\right)^2,
\end{equation}
where $\mu_m=\frac{1}{HW}\sum_{i=0}^{H-1}\sum_{j=0}^{W-1}\bigl[\mathbf{M}\left(\mathcal{E}^\prime(\bm{\theta})\right)\bigr]_{ij}$ is the mean of $\bigl[\mathbf{M}\left(\mathcal{E}^\prime(\bm{\theta})\right)\bigr]_{ij}$.
EA is then formulated as
\begin{equation}\label{eq:ea_formulation}
    \max_{\boldsymbol{\theta}} f_\text{EA}(\boldsymbol{\theta}).
\end{equation}

\subsubsection{ED Formulation}

For simplicity, we now consider ED on non-aligned observed events $\mathcal{E}=\{(\mathbf{x}_k,t_k,p_k)\}^{N_e}_{k=1}$.
To newly formulate ED using the contrast map, we introduce a confidence map $\mathbf{W}\in[0,1]^{H\times W}$ whose entry $w_{ij}$ quantifies how likely the events at the $(i,j)$th cell are signal events. Using $\mathbf{W}$, we define a weighted contrast map $\mathbf{M}(\mathcal{E},\mathbf{W})$ whose $(i,j)$th entry is given by
\begin{equation}\label{eq:masked_iwe}
    \bigl[\mathbf{M}(\mathcal{E},\mathbf{W})\bigr]_{ij} \coloneqq w_{ij}\sum_{k=1}^{N_e}\mathbb{I}\bigl(\mathbf{x}_k \in \Omega_{ij}\bigr).
\end{equation}
ED tends to decrease the overall contrast level of $\mathbf{M}(\mathcal{E},\mathbf{W})$ by removing noise.
To evaluate this, we use the variance of $\mathbf{M}(\mathcal{E},\mathbf{W})$ in the same manner as \eqref{eq:ea_objective}:
\begin{equation}\label{eq:ed_objective}
    f_{\text{ED}}(\mathcal{E},\mathbf{W}) = \frac{1}{HW}\sum_{i=0}^{H-1}\sum_{j=0}^{W-1}\left(\bigl[\mathbf{M}(\mathcal{E},\mathbf{W})\bigr]_{ij}-\mu_w\right)^2,
\end{equation}
where $\mu_w=\frac{1}{HW}\sum_{i=0}^{H-1}\sum_{j=0}^{W-1}\bigl[\mathbf{M}(\mathcal{E},\mathbf{W})\bigr]_{ij}$ is the mean of $\bigl[\mathbf{M}(\mathcal{E},\mathbf{W})\bigr]_{ij}$.
ED then can be treated as minimization of \eqref{eq:ed_objective} in the contrast domain:
\begin{equation}\label{eq:ed_formulation}
    \min_{\mathbf{W}}f_{\text{ED}}(\mathcal{E},\mathbf{W}).
\end{equation}

\subsection{Formulation as Regret-based Pareto Optimization}
A joint optimization of EA and ED consists of two conflicting objectives: EA seeks to increase the contrast in \eqref{eq:ea_formulation}, whereas ED acts to decrease it in \eqref{eq:ed_formulation}, as shown in Fig.~\ref{fig:idea_method}. 
In the following, we cast these conflicting problems as a bi-objective Pareto optimization problem \cite{coellocoelloComprehensiveSurveyEvolutionarybased1999,emmerichTutorialMultiobjectiveOptimization2018,konakMultiobjectiveOptimizationUsing2006,liManyObjectiveEvolutionaryAlgorithms2015,marlerSurveyMultiobjectiveOptimization2004,wangSurveySearchStrategy2023,zitzlerTutorialEvolutionaryMultiobjective2004}.

For joint optimization, ED should be evaluated on aligned events $\mathcal{E}^\prime(\bm{\theta})$. By substituting $\mathcal{E}$ with $\mathcal{E}^\prime(\bm{\theta})$ in \eqref{eq:ed_formulation}, we obtain $f_{\text{ED}}(\mathcal{E}^\prime(\boldsymbol{\theta}),\mathbf{W})$.
Let $\mathbf{f}(\bm{\theta},\mathbf{W})=\left[-f_{\text{EA}}(\bm{\theta}), f_{\text{ED}}\left(\mathcal{E}^\prime(\bm{\theta}),\mathbf{W}\right)\right]^\top\in\mathbb{R}^2$ be the vector valued-objective for joint EA and ED. The joint optimization problem for \eqref{eq:ea_formulation} and \eqref{eq:ed_formulation} can be written as follows:
\begin{equation}\label{eq:ea_ed_pareto}
    \min_{\boldsymbol{\theta},\mathbf{W}} \mathbf{f}(\boldsymbol{\theta},\mathbf{W}).
\end{equation}
Generally, the solution of \eqref{eq:ea_ed_pareto} is a set of optimal trade-offs between $f_{\text{EA}}$ and $f_{\text{ED}}$: No solution on this set can reduce $-f_{\text{EA}}$
without increasing $f_{\text{ED}}$, and vice versa.

To select a single solution, we utilize the Chebyshev minimax regret scalarization \cite{savage1951theory, wierzbicki1980use}. This aims to minimize the worst-case difference from a baseline value that represents an ideal performance level for each objective. 

Let $b_{\text{EA}}$ and $b_{\text{ED}}$ be the baselines for $f_{\text{EA}}$ and $f_{\text{ED}}$, respectively, where, $b_{\text{EA}}$ is empirically set and $b_{\text{ED}}$ is set to the variance of $\mathbf{M}(\mathcal{E})$. We define the regret with respect to each objective as follows:
\begin{equation}\label{eq:regrets}
    \begin{aligned}
        r_{\text{EA}}(\boldsymbol{\theta})
        &\coloneqq b_{\text{EA}} - f_{\text{EA}}(\boldsymbol{\theta})\\
        r_{\text{ED}}(\boldsymbol{\theta},\mathbf{W})
        &\coloneqq f_{\text{ED}}\left(\mathcal{E}^\prime(\bm{\theta}),\mathbf{W}\right) - b_{\text{ED}}.
    \end{aligned}
\end{equation}
A small regret means that the solution is close to its baseline. The worst-case regret over EA and ED is thus given by
\begin{equation}
    R(\bm{\theta},\mathbf{W})\coloneqq\max\left(r_{\text{EA}}(\boldsymbol{\theta}),r_{\text{ED}}(\boldsymbol{\theta},\mathbf{W})\right).
\end{equation}
Using this worst-case regret, we can transform the bi-objective problem \eqref{eq:ea_ed_pareto} into the following single-objective problem:
\begin{equation}\label{eq:max_regret_propose}
    \min_{\bm{\theta},\mathbf{W}}  R(\bm{\theta},\mathbf{W}).
\end{equation}

However, the problem \eqref{eq:max_regret_propose} admits trivial solutions (e.g., all entries in $\mathbf{W}$ to be $1$ or $0$). To avoid such solutions, we augment \eqref{eq:max_regret_propose} with the following two regularization terms: A sparsity penalty $\|\mathbf{W}\|_1$ reflecting the prior that signal events should concentrate on a small set of edge pixels, and a fidelity term $\|\mathbf{M}(\mathcal{E}^\prime(\boldsymbol{\theta}), \mathbf{W})-\mathbf{M}\left(\mathcal{E}^\prime(\boldsymbol{\theta})\right)\|_{\text{F}}^2$ suppressing the oversmoothing.
The final formulation is thus given by a combination of \eqref{eq:max_regret_propose} and regularization terms:
\begin{equation}\label{eq:final_MOP}
    \min_{\boldsymbol{\theta},\mathbf{W}}R(\bm{\theta},\mathbf{W})+\alpha\|\mathbf{W}\|_1+\beta\|\mathbf{M}(\mathcal{E}^\prime(\boldsymbol{\theta}), \mathbf{W})-\mathbf{M}\left(\mathcal{E}^\prime(\boldsymbol{\theta})\right)\|_{\text{F}}^2.
\end{equation}
Since \eqref{eq:final_MOP} is differentiable\footnote{In practice, $\delta$ in \eqref{eq:contrast_map} is typically relaxed to a Gaussian kernel, i.e., $\delta(\mathbf{x}-\bm{\mu})\approx\mathcal{N}(\mathbf{x}; \bm{\mu},\mathbf{I})$ to ensure differentiability.} with respect to both $\boldsymbol{\theta}$ and $\mathbf{W}$, it can be easily solved by applying the Adam algorithm \cite{kingma2014adam}.

\section{Experiments}\label{sec:experiments}

In this section, we evaluate our proposed method through denoising and motion estimation experiments.
Since there is no standard benchmark that simultaneously evaluates both ED and EA, we assess ED and EA separately. 

For ED, we conduct denoising experiments to compare the performance of the proposed method with a wide range of existing ED methods and the existing joint approach \cite{shibaSimultaneousMotionNoise2025} (abbreviated as EJA). 

For EA, we conduct motion estimation experiments. Since EA estimates MVs, EA performance is commonly assessed by comparing estimated MVs with the ground truth motion data from inertial measurement unit (IMU) sensors.

The evaluation consists of the following three experimental settings:
\begin{description}
    \item[Denoising on a Large-Scale Dataset:] 
    We first evaluate the proposed method on E-MLB dataset \cite{dingEMLBMultilevelBenchmark2023a}, which serves as a de facto benchmark for ED. This experiment compares the denoising performance of the proposed method against a wide range of existing ED methods.

    \item[Denoising Comparison with an Existing Joint Approach:] 
    We further conduct denoising experiments on DND21 dataset \cite{guoLowCostLatency2022}. To verify how the trade-off between EA and ED is controlled in formulation \eqref{eq:final_MOP}, we directly compare the denoising performance of the proposed method with EJA \cite{shibaSimultaneousMotionNoise2025}, introduced in \ref{subsec:joint}.
    \item[Motion Estimation:] 
    Finally, we assess the motion estimation performance of the proposed method on ECD dataset \cite{mueggler2017event}. We compare our method with a sequential approach (ED followed by EA) and the EJA method.
\end{description}

In the following, we describe these experiments.

\subsection{Denoising on a Large-Scale Dataset}\label{sub:denoising_experiment_emlb}
Here, we conduct denoising experiments on E-MLB dataset.

\subsubsection{Setup}

E-MLB dataset \cite{dingEMLBMultilevelBenchmark2023a} is a large-scale de facto dataset to benchmark ED performance. This dataset consists of $100$ sequences during both day and night, spanning from indoor to outdoor environments. For each sequence, event streams with four distinct lighting conditions (ND1, ND4, ND16, and ND64) are provided. The dataset was captured using a DAVIS346 \cite{taverni2018front}, which output event streams with $120$ dB at a resolution of $346\times260$.

We use event structural rate (ESR) \cite{dingEMLBMultilevelBenchmark2023a} to evaluate the denoising performance since E-MLB dataset does not provide event-wise ground truth annotations to distinguish noise events from signal ones. This metric measures how strongly events concentrate along object edges on the contrast map, instead of being dispersed as noise.
ESR is given by 
\begin{equation*}
    \text{ESR}=\sqrt{
      \begin{aligned}
        & \left(\sum_{i=0}^{H-1}\sum_{j=0}^{W-1}\frac{n_{ij}(n_{ij}-1)}{N_e(N_e-1)}\right) \\
        &\times \left(HW-\sum_{i=0}^{H-1}\sum_{j=0}^{W-1}(1-\frac{M}{N_e})^{n_{ij}}\right)
      \end{aligned}
    }\raisebox{-7ex}{,}
\end{equation*}
where $n_{ij}$ is the number of events that fall into the $(i,j)$th pixel on the contrast map, and $M$ is a reference number of events that compensates for differences in the number of denoised events across ED methods.
A higher ESR indicates better denoising performance, as denoised events are more concentrated along object edges. Further details can be found in \cite{dingEMLBMultilevelBenchmark2023a}.

    

We compare the denoising performance of the proposed method with seven model-based methods, including EJA \cite{delbruckFramefreeDynamicDigital2008, guoLowCostLatency2022, khodamoradiNspaceSpatiotemporalFilter2018, fengEventDensityBased2020a, lagorceHotsHierarchyEventbased2016, wangEVgaitEventbasedRobust2019, shibaSimultaneousMotionNoise2025}, and two learning-based methods \cite{baldwinEventProbabilityMask2020, guoLowCostLatency2022}. Unlike previous work \cite{dingEMLBMultilevelBenchmark2023a}, where the hyperparameters of each method are tuned separately for each sequence, we use fixed hyperparameters across all sequences in the dataset. 

\begin{figure*}[t]
  \centering

  \begin{subfigure}[t]{0.16\textwidth}
    \centering
    \includegraphics[width=\linewidth]{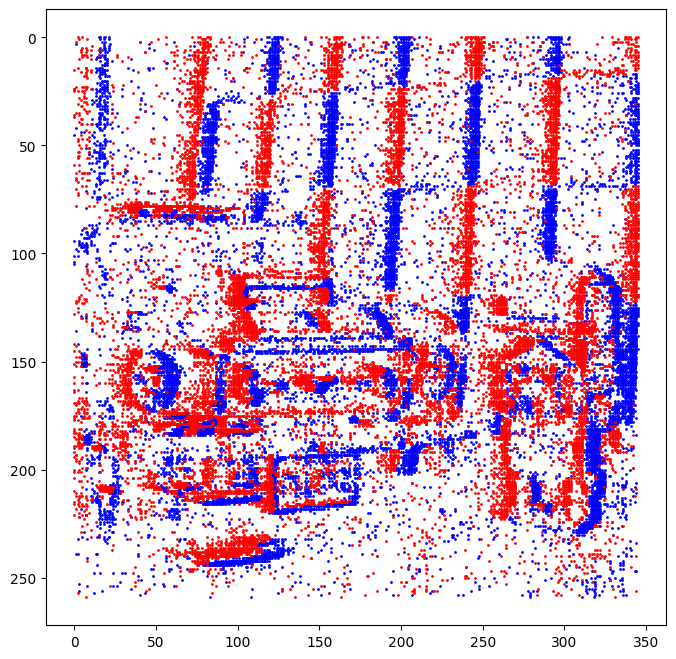}
    \caption{Raw Events}\label{fig:raw_emlb}
  \end{subfigure}\hfill
  \begin{subfigure}[t]{0.16\textwidth}
    \centering
    \includegraphics[width=\linewidth]{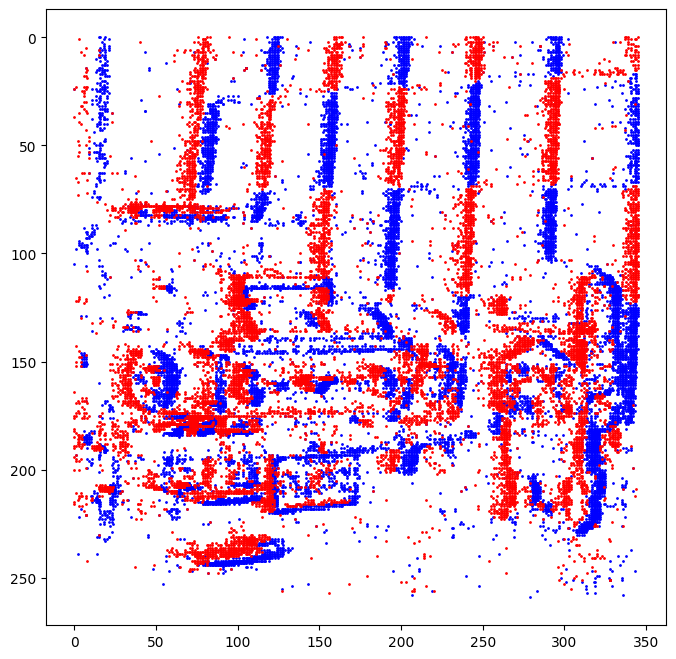}
    \caption{Denoised: BAF}\label{fig:baf_emlb}
  \end{subfigure}\hfill
  \begin{subfigure}[t]{0.16\textwidth}
    \centering
    \includegraphics[width=\linewidth]{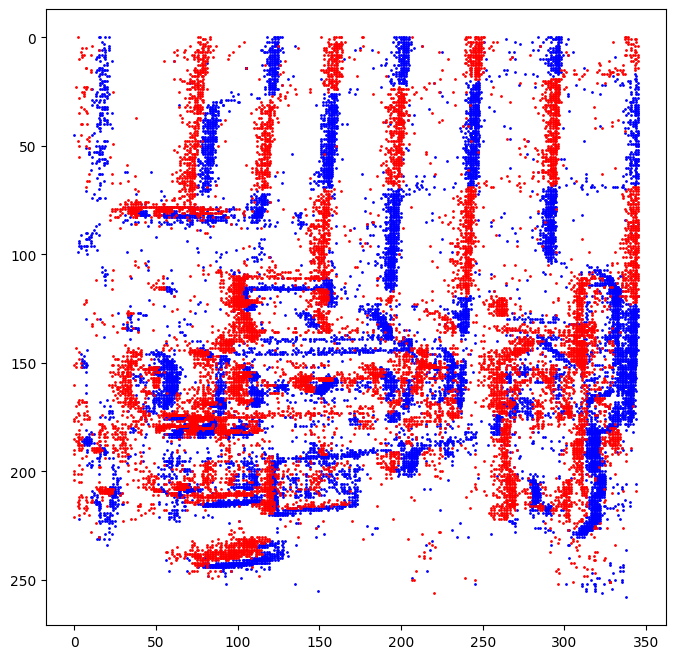}
    \caption{Denoised: DWF}\label{fig:dwf_emlb}
  \end{subfigure}
  \begin{subfigure}[t]{0.16\textwidth}
    \centering
    \includegraphics[width=\linewidth]{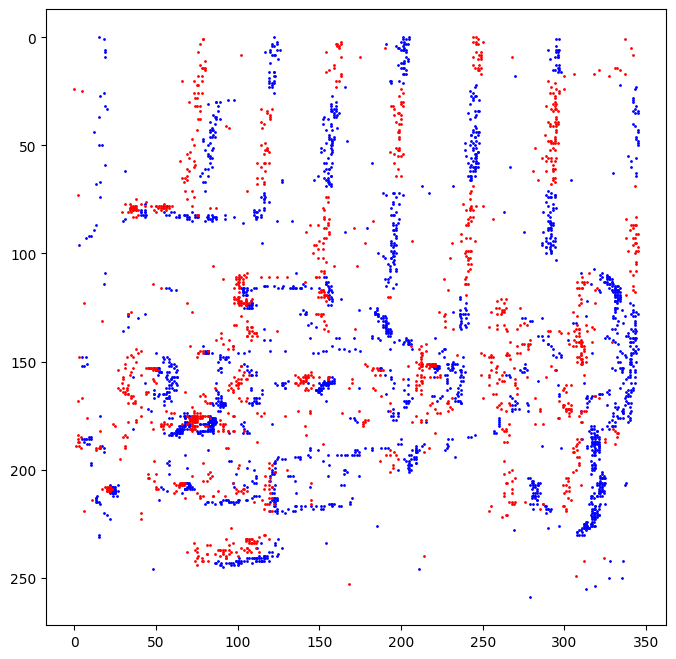}
    \caption{Denoised: KNoise}\label{fig:knoise_emlb}
  \end{subfigure}\hfill
  \begin{subfigure}[t]{0.16\textwidth}
    \centering
    \includegraphics[width=\linewidth]{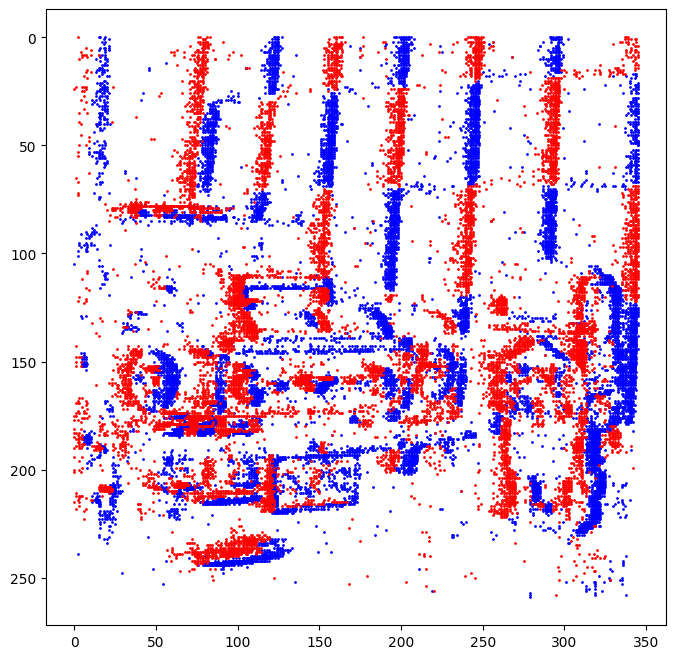}
    \caption{Denoised: YNoise}\label{fig:ynoise_emlb}
  \end{subfigure}\hfill
  \begin{subfigure}[t]{0.16\textwidth}
    \centering
    \includegraphics[width=\linewidth]{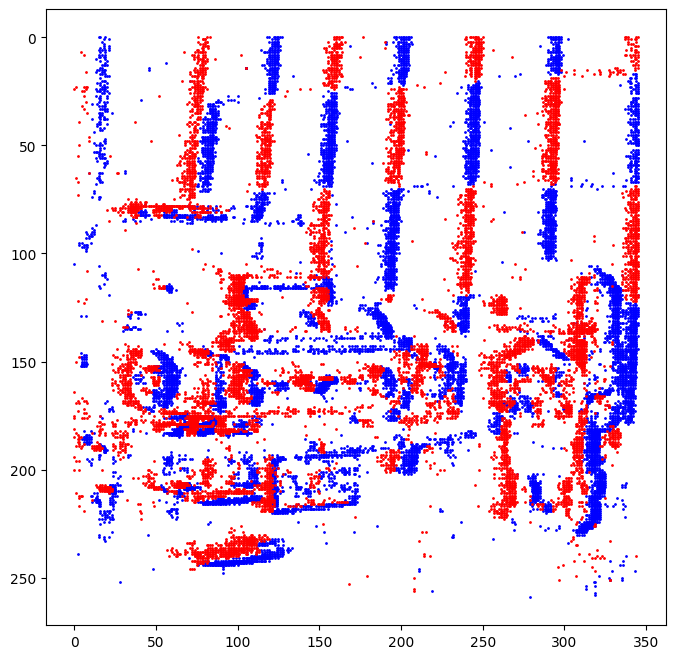}
    \caption{Denoised: TS}\label{fig:ts_emlb}
  \end{subfigure}

  \vspace{0.6em}

  \begin{subfigure}[t]{0.16\textwidth}
    \centering
    \includegraphics[width=\linewidth]{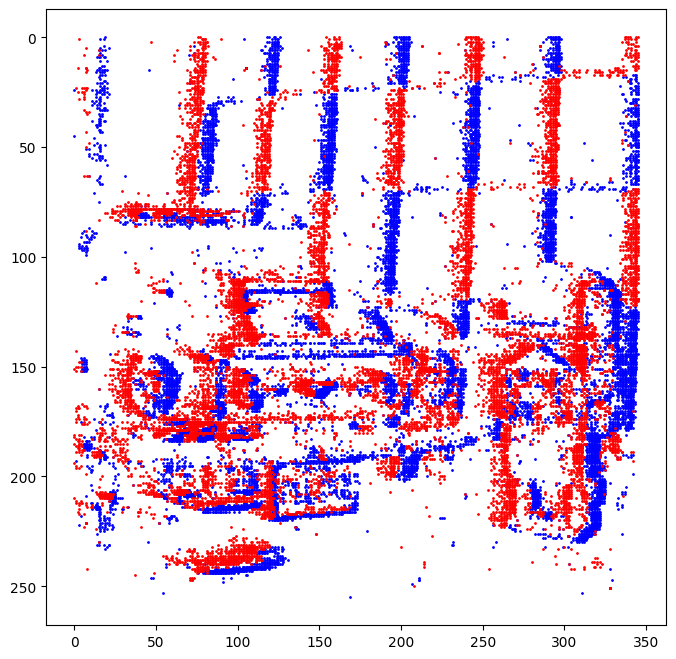}
    \caption{Denoised: EVFlow}\label{fig:evflow_emlb}
  \end{subfigure}
  \begin{subfigure}[t]{0.16\textwidth}
    \centering
    \includegraphics[width=\linewidth]{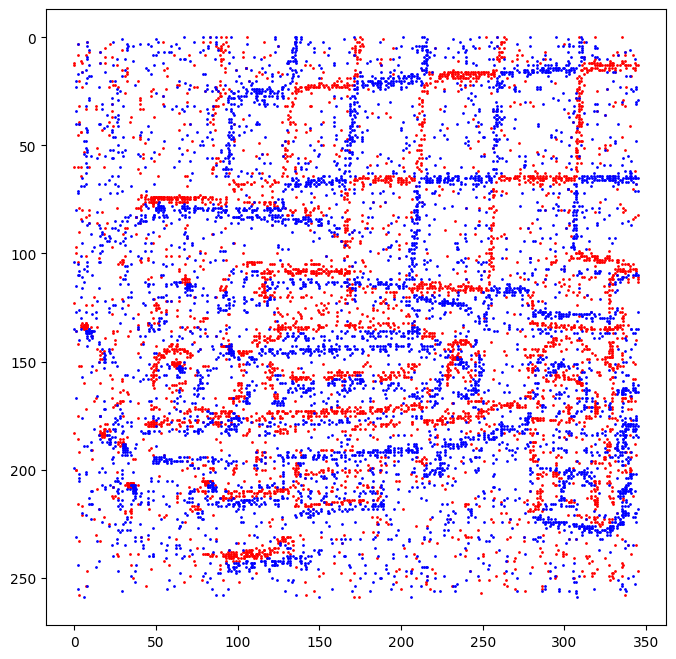}
    \caption{Denoised: EDnCNN}\label{fig:edn_emlb}
  \end{subfigure}
  \begin{subfigure}[t]{0.16\textwidth}
    \centering
    \includegraphics[width=\linewidth]{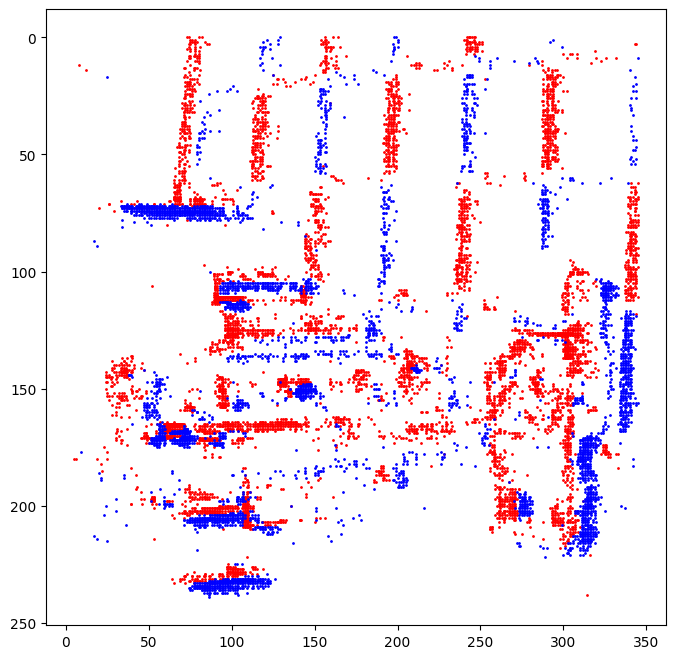}
    \caption{Denoised: MLPF}\label{fig:mlpf_emlb}
  \end{subfigure}
  \begin{subfigure}[t]{0.16\textwidth}
    \centering
    \includegraphics[width=\linewidth]{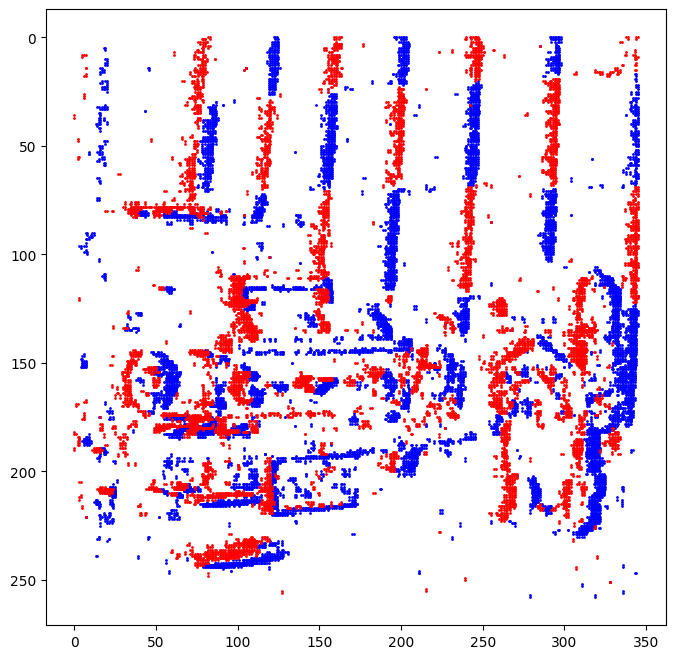}
    \caption{Denoised: EJA}\label{fig:EJA_emlb}
  \end{subfigure}
  \begin{subfigure}[t]{0.16\textwidth}
    \centering
    \includegraphics[width=\linewidth]{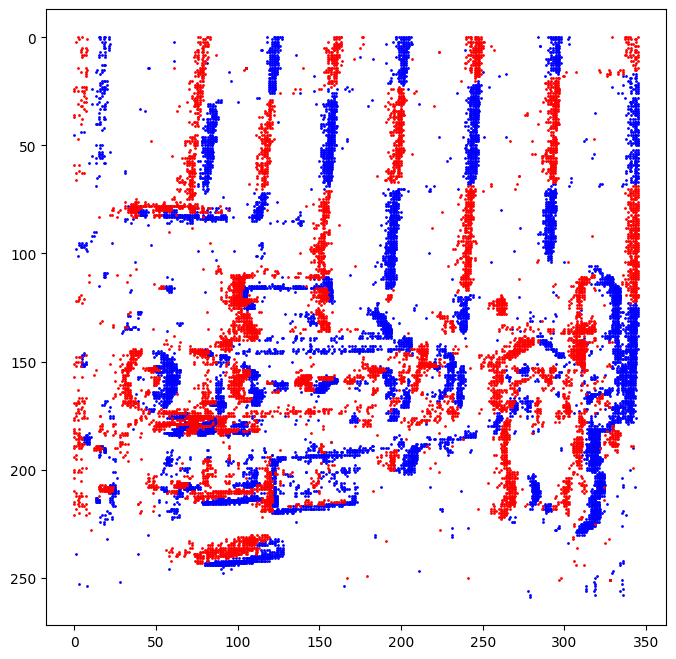}
    \caption{Denoised: Proposed}\label{fig:proposed_emlb}
  \end{subfigure}
  \hfill

  \caption[Scatter plots of denoised LEGO sequence in E\textendash MLB.]
  {Scatter plots of denoised LEGO sequence in E\textendash MLB dataset.
   Red points represent positive events; blue points represent negative events.}
  \label{fig:denoised_emlb}
\end{figure*}

\begin{table*}[tb]
\caption{ESR$\uparrow$ results on the E-MLB dataset. Bold numbers indicate the best results, while underlined numbers represent the second-best results.}
\centering
\small
\begin{tabular}{@{} l l l *{8}{c} c @{}}
\toprule
& & & \multicolumn{4}{c}{E-MLB (Day)} & \multicolumn{4}{c}{E-MLB (Night)} & \\
\cmidrule(lr){4-7} \cmidrule(lr){8-11}
Category & Type & Method & ND1 & ND4 & ND16 & ND64 & ND1 & ND4 & ND16 & ND64 & \#Hyperparams \\
\midrule
\midrule

\multirow{9}{*}{\makecell[l]{Model-\\based}}
& \multirow{7}{*}{\makecell[l]{ED\\only}}
& Raw
& $0.817$ & $0.823$ & $0.815$ & $0.785$
& $0.889$ & $0.822$ & $0.782$ & $0.761$
& $-$ \\

& & BAF \cite{delbruckFramefreeDynamicDigital2008}
& $0.851$ & $0.860$ & $0.865$ & $0.857$
& $0.977$ & $0.962$ & $0.911$ & $0.939$
& $2$ \\

& & DWF \cite{guoLowCostLatency2022}
& $0.886$ & $0.889$ & $0.890$ & $0.884$
& $1.017$ & $1.001$ & $0.955$ & $0.940$
& $3$ \\

& & KNoise \cite{khodamoradiNspaceSpatiotemporalFilter2018}
& $0.836$ & $0.827$ & $0.820$ & $0.808$
& $0.963$ & $0.956$ & $0.881$ & $0.853$
& $2$ \\

& & YNoise \cite{fengEventDensityBased2020a}
& $0.841$ & $0.847$ & $0.844$ & $0.842$
& $0.927$ & $0.918$ & $0.921$ & $0.985$
& $3$ \\

& & TS \cite{lagorceHotsHierarchyEventbased2016}
& $\textbf{0.919}$ & $0.903$ & $0.901$ & $0.882$
& $\textbf{1.042}$ & $\textbf{1.065}$ & $1.015$ & $1.003$
& $3$ \\

& & EVFlow \cite{wangEVgaitEventbasedRobust2019}
& $0.800$ & $0.785$ & $0.779$ & $0.731$
& $0.967$ & $0.968$ & $0.869$ & $0.831$
& $3$ \\

\cmidrule(l){2-12}

& \multirow{2}{*}{\makecell[l]{Joint\\approach}}
& EJA \cite{shibaSimultaneousMotionNoise2025}
& $0.875$ & $0.900$ & $0.902$ & $0.911$
& $0.982$ & $0.995$ & $1.036$ & $\underline{1.152}$
& $2$ \\

& & Proposed
& $0.890$ & $\underline{0.912}$ & $\textbf{0.921}$ & $\textbf{0.930}$
& $0.994$ & $1.009$ & $\textbf{1.083}$ & $\textbf{1.263}$
& $3$ \\

\midrule
\midrule

\multirow{2}{*}{\makecell[l]{Learning-\\based}}
& \multirow{2}{*}{\makecell[l]{ED\\only}}
& EDnCNN \cite{baldwinEventProbabilityMask2020}
& $0.676$ & $0.731$ & $0.731$ & $0.732$
& $0.827$ & $0.728$ & $0.669$ & $0.630$
& $-$ \\

& & MLPF \cite{guoLowCostLatency2022}
& $\underline{0.916}$ & $\textbf{0.925}$ & $\underline{0.921}$ & $\underline{0.919}$
& $\underline{1.028}$ & $\underline{1.011}$ & $\underline{1.053}$ & $1.120$
& $-$ \\

\bottomrule
\end{tabular}
\label{tab:denoising_results_emlb}
\end{table*}

\subsubsection{Results}

\begin{description}
    \item[Qualitative Results]
    Figure~\ref{fig:denoised_emlb} visualizes the examples of the denoised events. We obtain these visualizations by mapping denoised events in 2-D spatial coordinates.
    As observed in Fig.~\ref{fig:denoised_emlb}, the proposed method shows better signal-noise separation compared to the alternative methods. For example, KNoise and MLPF oversmooth event streams, resulting in faded edges (see Fig.~\ref{fig:knoise_emlb} and Fig.~\ref{fig:mlpf_emlb}), whereas other alternative methods retain many noise events (see Fig.~\ref{fig:baf_emlb} and Fig.~\ref{fig:dwf_emlb}).

    \item[Quantitative Results]
    ESRs are summarized in Table~\ref{tab:denoising_results_emlb}.
    We categorize the alternative methods into model-based and learning-based methods. 
    As shown in Table~\ref{tab:denoising_results_emlb}, the proposed method and EJA outperform other alternative methods in most cases.
    This implies that taking account of EA in ED improves ED performance. Notably, the proposed method outperforms EJA and often achieves the highest ESR. This may indicate that our regret-based formulation \eqref{eq:final_MOP} can balance the trade-off between EA and ED more robustly than EJA.
    
\end{description}

So far, we have demonstrated that the joint methods improve denoising performance by considering the trade-off between EA and ED. This motivates a direct comparison between the proposed method and EJA. In the following, we perform additional denoising experiments to clarify the differences between these two methods.

\subsection{Denoising Comparison with EJA}\label{sub:denoising_experiment_dnd}
We perform denoising experiments on a dataset with known noise rates to assess how each method controls the trade-off between EA and ED.
\subsubsection{Setup}
We use DND21 dataset \cite{guoLowCostLatency2022}. This dataset has been widely used for ED evaluation \cite{dingEMLBMultilevelBenchmark2023a, guoLowCostLatency2022, jiang2024edformer}. Since the recorded sequences are aggressively denoised by a denoising filter, they can be regarded as ``clean" signal event streams. We label all original events as signals and add randomly distributed events as noise to the original event streams.
We synthesize three event streams where the noise rates are set to 1\%, 5\%, and 10\% of the original data. The dataset was also captured using a DAVIS346 camera.

We use sensitivity and specificity to evaluate the denoising performance. 
Sensitivity represents the ratio of the number of detected signal events (TP: true positive) to that of entire signal events. Denoting the number of the undetected real events (FN: false negative), sensitivity is measured by $\text{Sensitivity} = \frac{\text{TP}}{\text{TP} + \text{FN}}$.
On the other hand, specificity represents the ratio of the number of detected noise events (TN: true negative) to that of entire noise events.
Denoting the number of the undetected noise events (FP: false positive), specificity is measured by $\text{Specificity} = \frac{\text{TN}}{\text{TN} + \text{FP}}$.

In the experiments, we use three variants for each method, where the parameters are tuned for typical noise rates (1\%, 5\%, 10\%), and evaluate every variant on all noise rate settings. 

\subsubsection{Results}
\begin{figure}[t]
  \centering

  \begin{subfigure}[t]{0.15\textwidth}
    \centering
    \includegraphics[width=\linewidth]{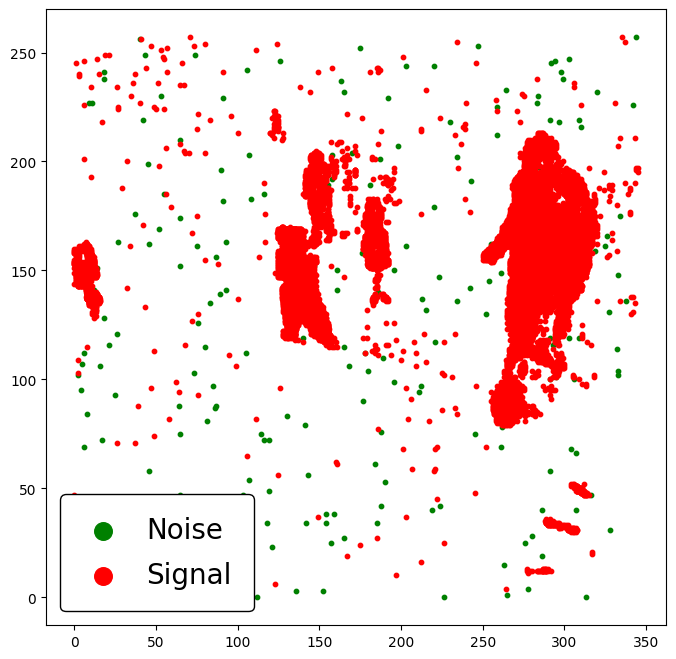}
    \caption{Raw Events w/ 1\% noise rate}\label{fig:raw1_dnd}
  \end{subfigure}\hfill
  \begin{subfigure}[t]{0.15\textwidth}
    \centering
    \includegraphics[width=\linewidth]{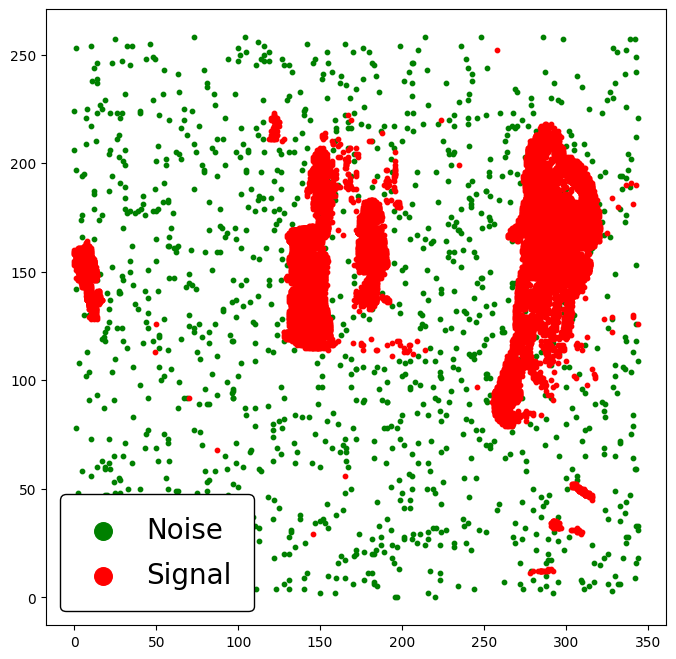}
    \caption{Raw Events w/ 5\% noise rate}\label{fig:raw5_dnd}
  \end{subfigure}\hfill
  \begin{subfigure}[t]{0.15\textwidth}
    \centering
    \includegraphics[width=\linewidth]{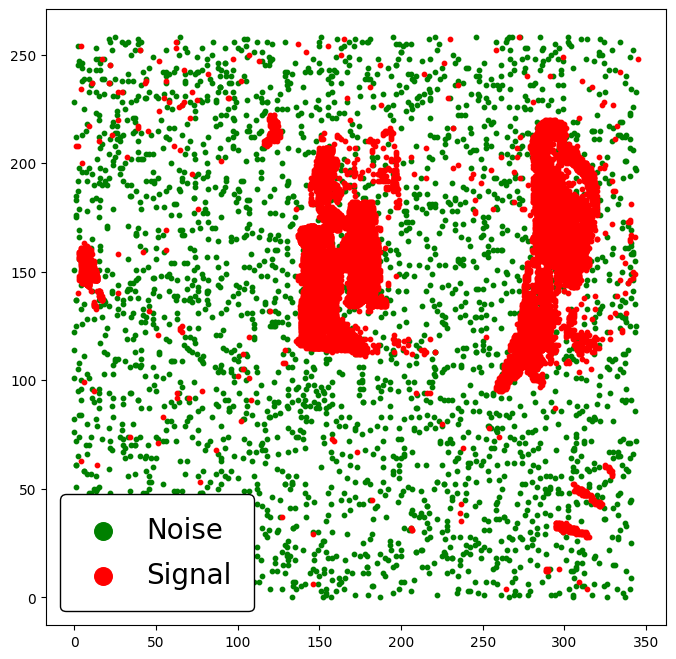}
    \caption{Raw Events w/ 10\% noise rate}\label{fig:raw10_dnd}
  \end{subfigure}

  \vspace{0.6em}
  
  \begin{subfigure}[t]{0.15\textwidth}
    \centering
    \includegraphics[width=\linewidth]{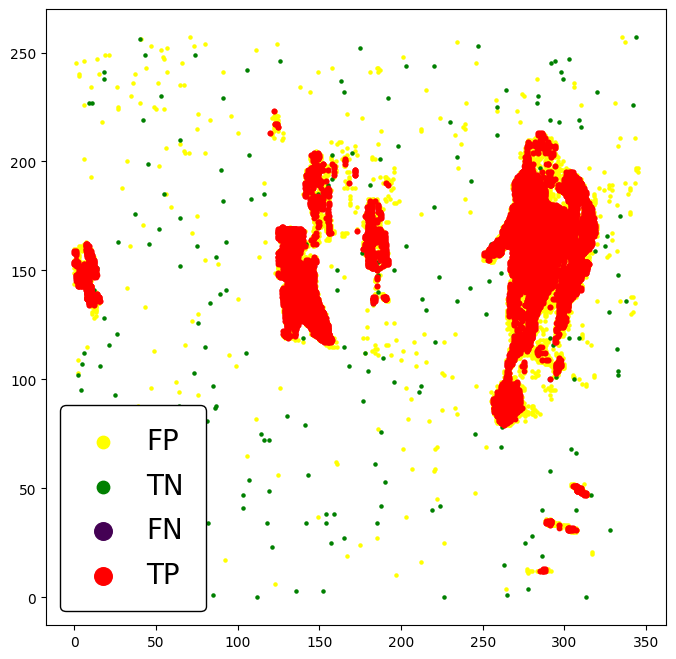}
    \caption{Denoised: EJA tuned for 1\% noise}\label{fig:EJA1_dnd}
  \end{subfigure}\hfill
  \begin{subfigure}[t]{0.15\textwidth}
    \centering
    \includegraphics[width=\linewidth]{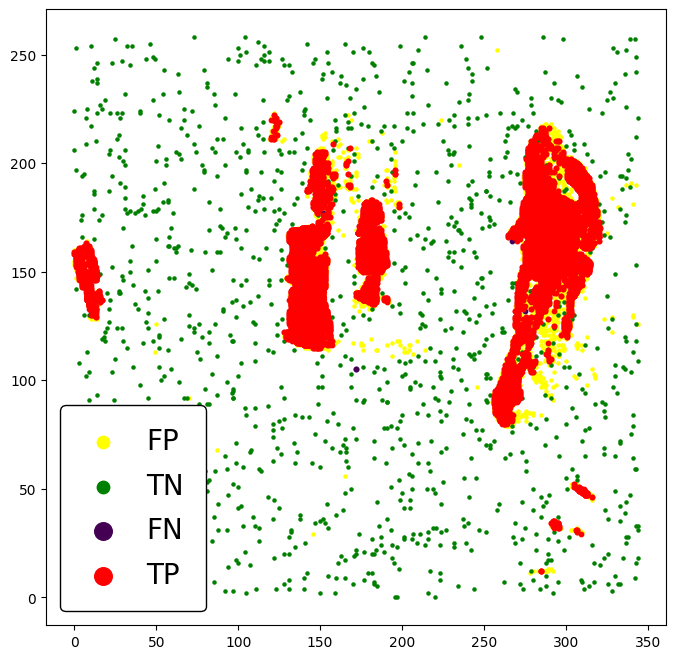}
    \caption{Denoised: EJA tuned for 5\% noise}\label{fig:EJA5_dnd}
  \end{subfigure}\hfill
  \begin{subfigure}[t]{0.15\textwidth}
    \centering
    \includegraphics[width=\linewidth]{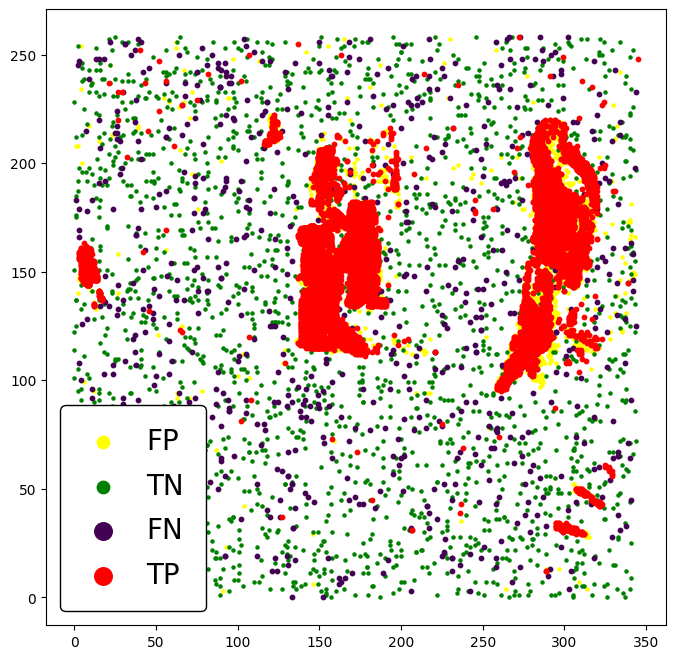}
    \caption{Denoised: EJA tuned for 10\% noise}\label{fig:EJA10_dnd}
  \end{subfigure}

  \vspace{0.6em}

  \begin{subfigure}[t]{0.15\textwidth}
    \centering
    \includegraphics[width=\linewidth]{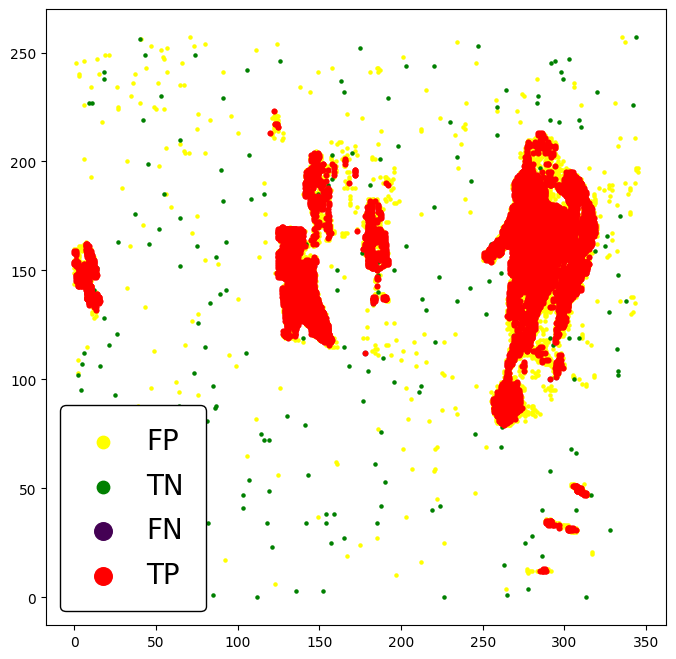}
    \caption{Denoised: Proposed tuned for 1\% noise}\label{fig:pro1_dnd}
  \end{subfigure}\hfill
  \begin{subfigure}[t]{0.15\textwidth}
    \centering
    \includegraphics[width=\linewidth]{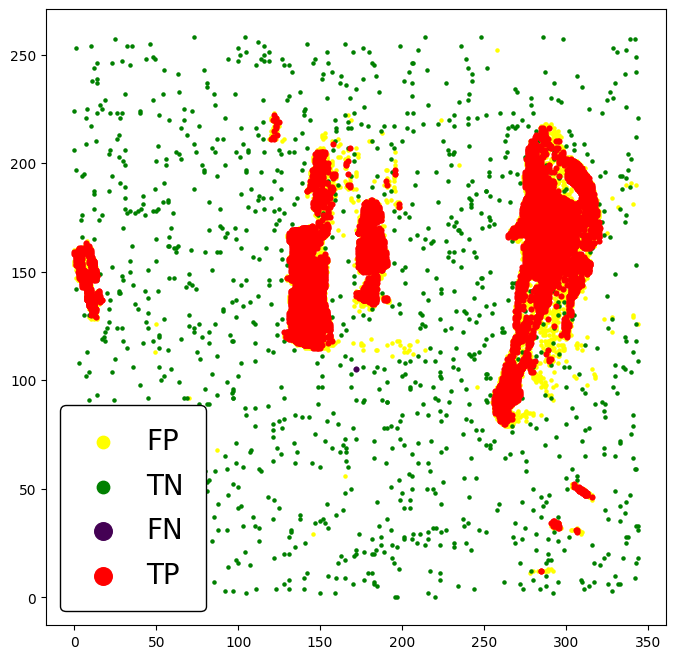}
    \caption{Denoised: Proposed tuned for 5\% noise}\label{fig:pro5_dnd}
  \end{subfigure}\hfill
  \begin{subfigure}[t]{0.15\textwidth}
    \centering
    \includegraphics[width=\linewidth]{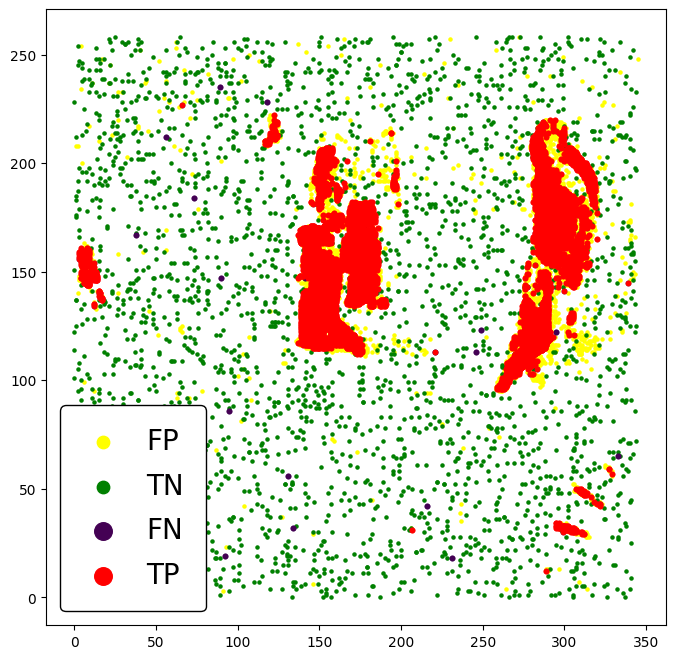}
    \caption{Denoised: Proposed tuned for 10\% noise}\label{fig:pro10_dnd}
  \end{subfigure}

  \caption[Scatter plots of denoised \textit{hotel\_bar} in DND21 dataset.]
  {Scatter plots of denoised \textit{hotel\_bar} in DND21 dataset.
   Red points: detected signal events (TP). Yellow points: undetected signal events (FP). Green points: detected noise events (TN). Black points: undetected noise events (FN).}
  \label{fig:denoised_dnd}
\end{figure}

\begin{description}
    \item[Qualitative Results]
    Examples of the denoised events are visualized in Fig.~\ref{fig:denoised_dnd}. 
    As observed in Fig.~\ref{fig:denoised_dnd}, the performance of EJA is sensitive to its tuning; it shows low FPs and FNs when the parameters of EJA are precisely matched to the noise rate of the data. However, its performance degrades when the parameter is not correctly tuned for the data. For instance, a large number of noise events remain undetected when the noise rate is underestimated (see Fig.~\ref{fig:EJA10_dnd}). In contrast, our proposed method consistently demonstrates low FPs and FNs across various noise levels.
    
    \item[Quantitative Results]
    The experimental results on DND21 are summarized in Table~\ref{tab:denoising_dnd}.
    According to the table, EJA achieves good performance when its tuned parameter perfectly matches the ground-truth noise rate (see the diagonal entries). 
    However, a mismatch between the noise rate and the tuned parameter produces a see-saw effect: over-tuning to higher noise increases TNR but lowers TPR (signal loss), while under-tuning does the opposite (noise leakage).
    This indicates that EJA falls into a suboptimal solution when the correct noise rate is not provided.
    In contrast, the proposed method maintains consistently high and stable performance in both TPR and TNR across all noise levels. This indicates that regret-based formulation \eqref{eq:final_MOP} can control the trade-off between EA and ED robustly, against varying noise rates.
\end{description}

\begin{table}[tb]
\caption{Sensitivity and Specificity results on DND21 dataset. Bold numbers indicate the best results while underlined numbers represent the second-best results.}
  \centering
  \small 
  \setlength{\tabcolsep}{4pt} 
  \begin{tabular}{@{}ll ccc ccc@{}} 
    \toprule
    & & \multicolumn{3}{c}{Sensitivity} & \multicolumn{3}{c}{Specificity} \\
    \cmidrule(lr){3-5} \cmidrule(lr){6-8}
    & noise rate (\%) & 1\% & 5\% & 10\% & 1\% & 5\% & 10\% \\
    \midrule
    \midrule
    & \makecell[l]{EJA tuned for 1\%} & \textbf{0.918} &\textbf{0.945} & \textbf{0.962} & 0.936 & 0.787 & 0.609 \\
    & \makecell[l]{EJA tuned  for 5\%} & 0.888 & 0.920 & \underline{0.944} & \underline{0.956} & 0.918 & 0.760 \\
    & \makecell[l]{EJA tuned for 10\%} & 0.838 & 0.869 & 0.905 & \textbf{0.963} & \textbf{0.955} & 0.920 \\ 
    \midrule
    \midrule
    & \makecell[l]{Proposed tuned for 1\%} & \underline{0.917} & 0.915 & 0.913 & 0.950 & 0.945 & \underline{0.939} \\
    & \makecell[l]{Proposed tuned for 5\%} & 0.911 & \underline{0.921} & 0.915 & 0.937 & \underline{0.951} & 0.936 \\
    & \makecell[l]{Proposed tuned for 10\%} & 0.908 & 0.913 & 0.918 & 0.933 & 0.942 & \textbf{0.944} \\
    \bottomrule
  \end{tabular}
  \label{tab:denoising_dnd}
\end{table}

\subsection{Motion Estimation Experiments}\label{sub:motion_experiments}
Here, we conduct motion estimation experiments.

\subsubsection{Setup}
We perform motion estimation experiments on ECD dataset \cite{mueggler2017event}. This dataset is a standard dataset for evaluating various tasks like camera ego-motion estimation.
Each sequence in this dataset provides events ($130$ dB), frames, calibration information, IMU data, and ground truth (GT) camera poses (at $200$ Hz). We use \textit{dynamic\_rotation}, \textit{boxes\_rotation}, \textit{dynamic\_translation}, and \textit{boxes\_translation} sequences to evaluate motion estimation accuracy.
The data was captured using a DAVIS240C at a resolution of $240\times 180$.

To quantitatively evaluate the accuracy of our motion estimation, we use Root Mean Square Error (RMSE). This metric measures the closeness between the estimated motion parameters and the ground truth IMU data. The RMSE  is computed by
\begin{equation*}
 \text{RMSE} = \sqrt{\frac{1}{T} \sum_{i=1}^{T} \| \boldsymbol{\theta}_{\text{est}}(t_i) - {\boldsymbol{\theta}}_{\text{gt}}(t_i) \|^2_2},
\end{equation*}
where $\boldsymbol{\theta}_{\text{est}}(t_i)$ is the set of estimated motion parameters at $t_i$ and $\boldsymbol{\theta}_{\text{gt}}(t_i)$ is the corresponding ground truth IMU data, respectively.

 We compare the motion estimation accuracy of the proposed method with the following two approaches: 1) sequential implementation of a denoising filter \cite{delbruckFramefreeDynamicDigital2008} and Cmax \cite{gallegoUnifyingContrastMaximization2018,shiba2024secrets}, and 2) EJA.

\subsubsection{Results}

\begin{figure}[t]
  \centering
  \begin{subfigure}[t]{0.15\textwidth}
    \centering
    \includegraphics[width=\linewidth]{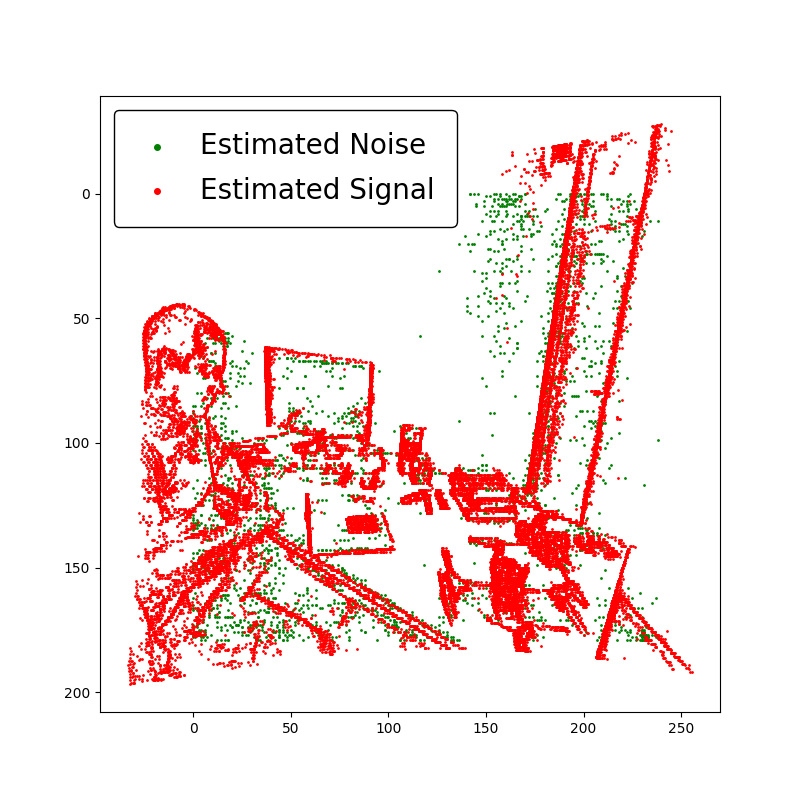}
    \caption{BAF \& CMax}\label{fig:baf_ecd}
  \end{subfigure}\hfill
  \begin{subfigure}[t]{0.15\textwidth}
    \centering
    \includegraphics[width=\linewidth]{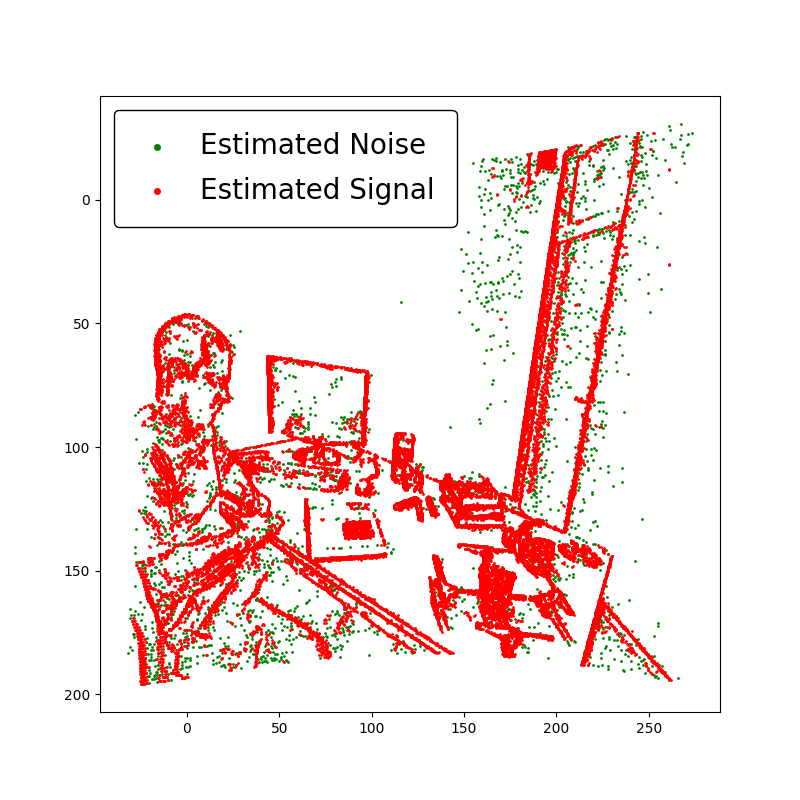}
    \caption{EJA}\label{fig:sim_ecd}
  \end{subfigure}\hfill
  \begin{subfigure}[t]{0.15\textwidth}
    \centering
    \includegraphics[width=\linewidth]{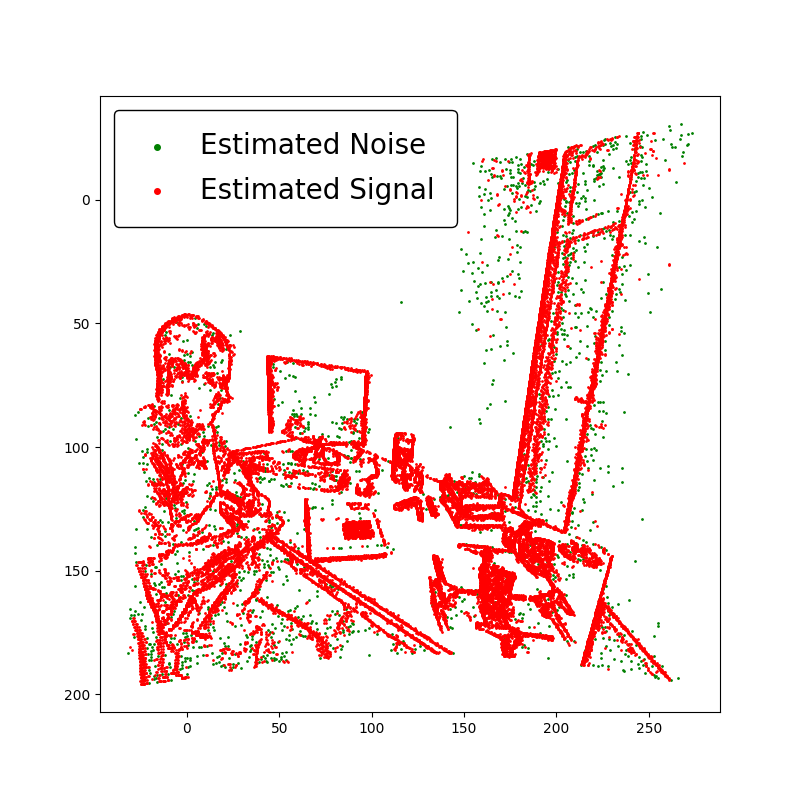}
    \caption{Proposed}\label{fig:proposed_ecd}
  \end{subfigure}
  \caption[Scatter plots of aligned \textit{dynamic\_rotation} in ECD dataset.]
  {Scatter plots of aligned \textit{dynamic\_rotation} in ECD dataset.
   Red points represent detected signal events ,while green points represent detected noise events.}
  \label{fig:alignment_ecd}
\end{figure}

\begin{description}
    \item[Qualitative Results] 
    We visualize the 2-D image of aligned events in Fig.~\ref{fig:alignment_ecd}. 
    As observed in Fig.~\ref{fig:alignment_ecd}, the proposed method and EJA seem to be aligned well compared to the sequential approach.
    
    \item[Quantitative Results] 
    RMSEs are summarized in Table~\ref{tab:motion_estimation_ECD}.
    According to the table, the two joint methods consistently outperform the sequential approach. This suggests that explicitly accounting for the presence of noise during alignment improves motion estimation accuracy. It is noteworthy that the proposed method shows the lowest RMSE on all sequences. This may indicate that joint optimization of EA and ED alleviates the chicken-and-egg dilemma.
\end{description}

\begin{table}[tb]
\caption{RMSE$\downarrow$ on ECD dataset. Bold numbers indicate the best results while underlined numbers represent the second-best results.}
  \centering

  \begin{tabular}{@{}llcccc@{}}
    \toprule
    & & \multicolumn{2}{c}{Rotation} & \multicolumn{2}{c}{Translation} \\
    \cmidrule(lr){3-4} \cmidrule(lr){5-6}
    & &
      \shortstack{\textit{dynamic}\\\textit{rotation}} &
      \shortstack{\textit{boxes}\\\textit{rotation}} &
      \shortstack{\textit{dynamic}\\\textit{translation}} &
      \shortstack{\textit{boxes}\\\textit{translation}} \\
    \midrule
    & BAF         & $6.973$                 & $68.673$ & $11.241$ & $27.228$ \\
    & EJA  & $\underline{6.660}$     & $\underline{64.167}$ & $\underline{10.517}$ & $\underline{26.939}$ \\
    & Proposed    & $\textbf{6.627}$        & $\textbf{56.364}$ & $\textbf{10.509}$ & $\textbf{26.661}$ \\
    \bottomrule
  \end{tabular}
  \label{tab:motion_estimation_ECD}
\end{table}

\section{Conclusion}\label{sec:conclusion}
This paper proposes a joint alignment and denoising method for EVSs.
To formulate EA and ED, we utilize the contrast map. On the contrast map, EA seeks to increase the contrast, whereas ED tends to reduce the overall contrast level. In the proposed method, we formulate these conflicting problems as a bi-objective Pareto optimization. This formulation provides a set of trade-off solutions between EA and ED. To control this trade-off and obtain a well-compromised solution, we use the regret strategy. 
In experiments, we demonstrate that the proposed method improves the performance of EA and ED by optimizing them jointly.

\bibliographystyle{IEEEtran}
\bibliography{bib}
\end{document}